# Band engineering of ternary metal nitride system Ti$_{1-x}$Zr$_x$N for plasmonic applications

**Mukesh Kumar,**[1,2,*] **Satoshi Ishii,**[2,3] **Naoto Umezawa,**[1,2] **and Tadaaki Nagao**[2,3]

[1]*Environmental Remediation Materials Unit, National Institute for Materials Science, Ibaraki 305-0044, Japan*
[2]*CREST, Japan Science and Technology Agency, 4-1-8 Honcho, Kawaguchi, Saitama, 332-0012, Japan*
[3]*International Center for Materials Nanoarchitectonics (MANA), National Institute for Materials Science (NIMS), Tsukuba 305-0044, Japan*

[*]*mkgarg79@gmail.com, Kumar.Mukesh@nims.go.jp*

**Abstract:** Chemical composition is the primary factor that determines the electronic band structure and thus also influences the optical properties of plasmonic ceramics including nitrides and oxides. In this work, the optical and plasmonic properties of TiN, ZrN and their hypothetical intermediate alloys Ti$_{1-x}$Zr$_x$N (x= 0, 0.25, 0.50, 0.75, and 1), are studied by using first-principles density functional theory. We demonstrate the effects of electronic band structure tuning (band engineering) on the dielectric properties by varying the concentration of metallic constituents. Our calculations reveal that bulk plasma frequency, onset of interband transitions, width of bulk plasmon resonance and cross-over frequency, can be tuned flexibly in visible spectrum region by varying the amount of Zr concentration in Ti$_{1-x}$Zr$_x$N alloy system. We found that low threshold interband energy onset (~1.95 eV) leads to high losses in Ti rich compounds than that of ZrN which points to lower losses.

## 1. Introduction

Plasmonics has opened the door for many technological applications such as bio-sensing [1], solar energy harvesting [2], plasmonic circuit [3] to name but a few in the past decade. Gold (Au) and silver (Ag) have been commonly used in these applications due to their large plasma frequencies and low losses (small plasmon damping). However, noble metals are too expensive for mass productions and the magnitudes of the real parts are too large, which is not suitable for fabricating metamaterials and transformation optics devices for practical uses [4]. Also the mechanical strength, as well as stability at high temperature are lacking in these noble metal elements. Because of these reasons, studies on alternative plasmonic materials such as intermetallics, metals alloys, transparent conductive oxides, and refractory compounds, have been on high demands in the past few years [5-10]. In this context, transition metal nitrides such as titanium nitride (TiN) and zirconium nitride (ZrN) were proposed for plasmonic applications in the visible frequencies range as alternative plasmonic materials [11-14]. Obviously they are much more inexpensive than noble metals and the real parts of permittivties are negative with reasonably large magnitudes. In addition, properties like high melting point temperature, chemical stability, and resistivity against abrasion, make them promising for the use in plasmonic applications in harsh environment [15-19]. Particularly TiN has been proposed for exciting applications such as hyberbolic metamaterials and plasmonic interconnects [19]. Moreover, due to their simple cubic lattice structures, these materials can be grown epitaxially on various substrates like Si, MgO and crystalline sapphire [6,17]. There are many reports where optical properties of TiN and ZrN for plasmonic application have been discussed, however, to the best of our knowledge not a single study is reported on alloy composition of Ti$_{1-x}$Zr$_x$N nor the tunability of their optical properties. With this regards, theoretical approach based on first-principles density functional theory (DFT), is quite powerful in the study of electronic and optical properties of materials due to its superiority in control. Therefore, present study is design accordingly and the electronic and optical properties of Ti$_{1-x}$Zr$_x$N (x= 0, 0.25, 0.50, 0.75 and 1) alloy system are studies within theoretical domain by using both DFT and electromagnetic (or Mie theory) simulations. Our study reveal that optical properties of hypothetical ternary nitride alloys Ti$_{1-x}$Zr$_x$N, vary smoothly between those of TiN and ZrN. With the help of band engineering, the optical parameters including cross-over frequency, bulk plasmon energy, and interband transitions can be tuned to required frequencies in the visible spectrum for desired plasmonic applications. Details are discussed in this article.

## 2. Model, methods and computational details

Electronic band structure, optical and plasmonic properties of materials can be obtained easily from number of various methods like time-dependent DFT, many-body perturbation theory and independent particle approximation (IPA) with appropriate relativistic correction [20-27]. Here, we apply IPA, a simplest and successful approach used previously to examine the dielectric function of many metallic systems [25-27]. Since the electron-hole interaction is well screened in metals hence for the materials studied here, the IPA is expected to work reasonably well. The local field effects, excitonic effects, and spin-orbit coupling are excluded in present study. Therefore, an error of ±0.5 eV in the energy levels are expected to some extent.

However, one should also note that the final effect of these energy levels on optical properties is not straight as in case of semiconductors. It is true that unlike semiconductors, GW corrections do not act as a rigid shift of the whole occupied band structure with respect to empty conduction bands. This is the reason that GW corrections are highly nontrivial in metals [25].

The dielectric function of metals have contribution from both intraband (intra) and interband (inter) electron transitions. Hence, the total dielectric function is the sum of both intra and inter terms,

$$\varepsilon(\omega) = \varepsilon_{intra}(\omega) + \varepsilon_{inter}(\omega) \qquad (1)$$

The intra part, $\varepsilon_{intra}(\omega)$ of dielectric function was obtained from simple Drude model scheme [25],

$$\varepsilon_{intra}(\omega) = 1 - \frac{\omega_p^2}{\omega(\omega + i\gamma)} \qquad (2)$$

where $\omega_p$ and $\gamma$ are the plasma frequency and damping parameter due to the dispersion of the electrons, respectively. One can obtained $\omega_p$ from electronic band structure as discussed elsewhere [26], whereas $\gamma$ can be either obtained from experimental data or from higher order calculation [28]. The inter part of dielectric function $\varepsilon_{inter}(\omega) = \varepsilon_1(\omega) + i\varepsilon_2(\omega)$ (where $\varepsilon_1$ is the real part and $\varepsilon_2$ is the imaginary part) was obtained from electronic band structure calculation. First, imaginary part was calculated by evaluating direct electronic transition between occupied and unoccupied electronic states and then real part was obtained from the Kramers-Kronig transformation as per method discussed elsewhere [29] and also in our previous studies [30, 31], where the dielectric function of various semiconductors were calculated.

Once dielectric function is evaluated, the electron loss function ELF($\omega$) and reflectivity R($\omega$) were evaluated from the following formulations, [28]

$$ELF(\omega) = \text{Im}\left(-\frac{1}{\varepsilon(\omega)}\right) \qquad (3)$$

$$R(\omega) = \left|\frac{(1-\sqrt{\varepsilon(\omega)})}{(1+\sqrt{\varepsilon(\omega)})}\right|^2 \qquad (4)$$

All these calculations of $Ti_{1-x}Zr_xN$ alloy system were performed using DFT as implemented in Vienna ab initio Simulation Package code [32]. Projector augmented wave-function type pseudo-potentials along with generalized gradient approximation of Perdew-Burke-Ernzerhof (PBE) [33] was used for the exchange-correlation potential. A Brillouin integration grid was built using the Monkhorst-pack scheme with 21×21×21 and 29×29×29 points for electronic band structure and optical permittivity calculation, respectively.

To model the alloy compositions, rocksalt structure (Fm-3m: space group No. 225) of TiN and ZrN was considered. Two atoms primitive cell, shown in Fig. 1(a), was used for the electronic structure calculations of base compound like TiN and ZrN, whereas 8 atoms supercell, shown in Fig. 1(b), was used for optical properties calculations of alloy composition of $Ti_{1-x}Zr_xN$. Here we considered only three compositions i.e. x=0.25, x=0.50, and x=0.75. However, one can study various other compositions as well by modeling a big supercell. For the geometrical optimization of alloy system $Ti_{1-x}Zr_xN$, we considered various configurations of the Ti, Zr atoms and relax the structure for both volume and ions convergence. The total energy calculation shows that the most stable configuration involves structures where Ti atoms occupy corner position than Zr atom as shown in Fig. 1(c). The difference in total energy was only 0.24 meV/atom. Once corner position was fixed for Ti atom, the Zr substitution was studies on each available lattice sites and energetically favorable position was considered for optical properties calculation. We also found that clustering of Ti or Zr atoms yields a higher formation enthalpy, and we therefore consider only dispersed distribution of these atoms. Only ordered alloy compounds were simulated as calculations require a periodic structure. We believe that studied alloy compositions keep the identical crystal configuration due to the similarity in atomic sizes of Ti and Zr. Since, the main aim of our study is to demonstrate the tunability of the optical properties of the alloy system therefore study on crystal stability was not considered which however is beyond the scope of current work. Moreover, it is also reported earlier that 8 atoms cell can describe various alloy compositions efficiently for alloy structure like $Au_xPt_{1-x}Al_2$ and the effect of disordered filling of cation metal sites on optical properties is relatively subtle [34].

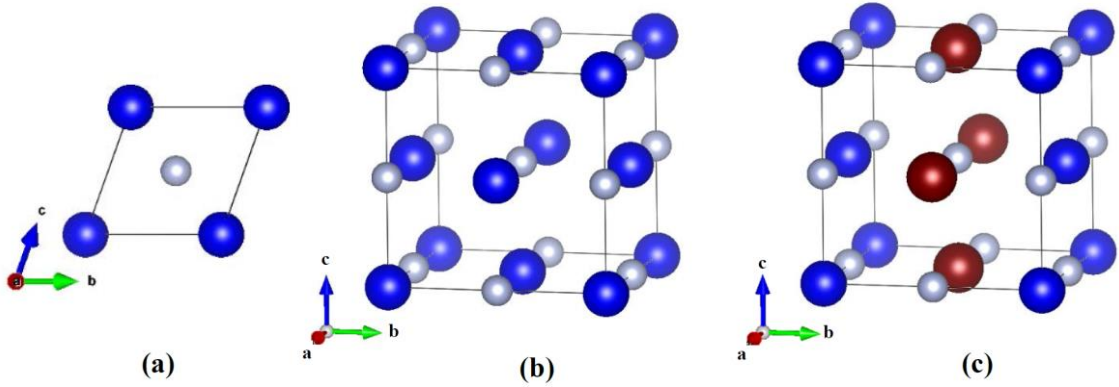

Fig.1 Crystal structure of the ordered alloy compounds (a) 2 atoms primitive cell of TiN, (b) 8 atoms cell of TiN and (c) 8 atoms cell of $Ti_{0.5}Zr_{0.5}N$ alloy. Here small grey balls represent N atom, whereas big blue and red ball represent Ti and Zr atoms, respectively.

## 3. Results

In this section we discuss the electronic and optical properties of $Ti_{1-x}Zr_xN$ alloy system. The calculated lattice parameters (TiN: 4.239 Å; $Ti_{0.75}Zr_{0.25}N$: 4.332 Å; $Ti_{0.5}Zr_{0.5}N$: 4.433 Å; $Ti_{0.25}Zr_{0.75}N$: 4.510 Å; and ZrN: 4.578 Å), are in good agreement with reported experimental values [35]. The linear increase in volume with Zr alloying shows the isostructural feature of these alloys.

### 3.1 Electronic properties

Figures 2(a) and 2(b) show the calculated electronic band structure of base compounds TiN and ZrN respectively.

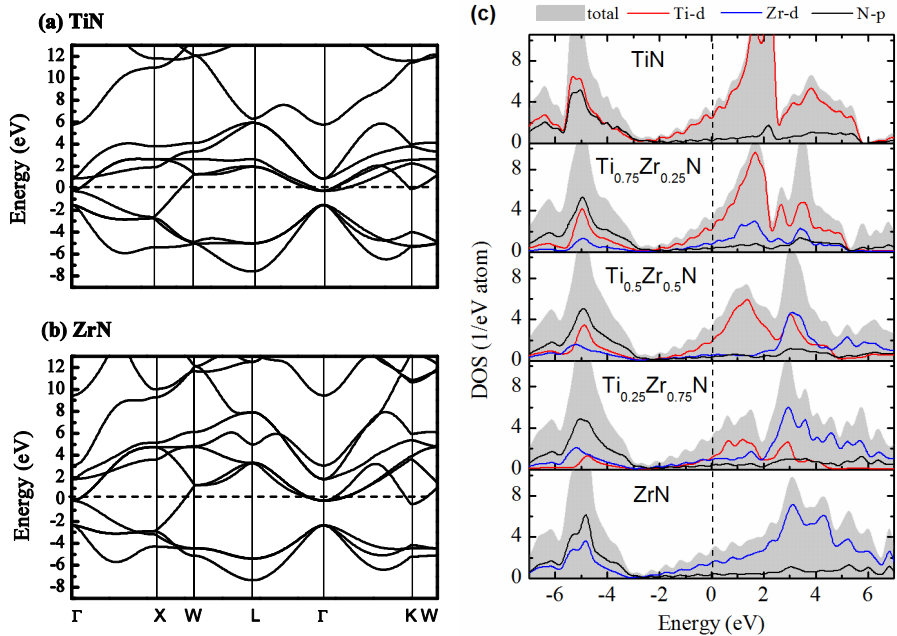

Fig.2 Calculated electronic band structure of (a) TiN, (b) ZrN along high symmetry directions and (c) DOS of $Ti_{1-x}Zr_xN$ alloy compounds. The dotted horizontal and vertical lines indicate the Fermi-level in left and right side panels, respectively.

Our calculated band structure of TiN and ZrN are consistent with earlier calculation [27]. One can clearly see that the bands cross the Fermi-level ($E_f$) (dotted horizontal lines) and hence no energy gap, confirms the metallic behavior of these materials. With the increase of lattice parameters from TiN → ZrN, the Fermi-level shifted upwards from the valence band (VB) region to the conduction band (CB) region. This can be attributed to the bonding/antibonding split off phenomena, as Ti $d$ states are lower in energy than Zr $d$ states. The basic difference comes from the band at Γ-point. This can be understood from the density of states (DOS) plots of these compounds. Figure 2(c) shows the DOS of $Ti_{1-x}Zr_xN$ alloy system. One can see that partially filled $d$ bands are located near the $E_f$. Since selection rule forbids the $d$-$d$ transitions, the only allowed optical transitions from $p \rightarrow d$, which is from hybridized N $p$ and Ti/Zr $d$ orbits along the Γ point of the 1$^{st}$ Brillouin zone. It is worth to mention that the position of the $p$ band with respect to the $E_f$ is lower in ZrN than in TiN by around 0.5 eV. This indicates that a higher energy is required for the interband $p \rightarrow d$ transitions to occur in ZrN. Based on the position of the band edge energies at the Γ point, one can estimate the threshold energy of the interband transitions which is related to the optical loss in materials. Our electronic structure calculations show that TiN has lower threshold energy (~ 1.9 eV), whereas ZrN has higher threshold energy (~2.6 eV). Hence, in the low energy region, it is expected that TiN has more optical losses than ZrN, whereas the alloy system lies between them.

*3.2    Optical properties*

Before discussing the alloy composition, we calculate the optical properties of Au, a bench mark material in plasmonic field, by using the same simulation scheme as described above. The calculated optical properties (shown in Table 1) shows that the predicted optical properties are in reasonable agreement with reported experimental results. Figure 3(a) and 3(b) show the calculated imaginary (optical losses) and real permittivity, respectively in Au along with the individual contribution from intraband and interband transitions. For comparison we also show the latest experiment data of single crystal Au sample [36] in the available range from 0.05 – 4.14 eV.

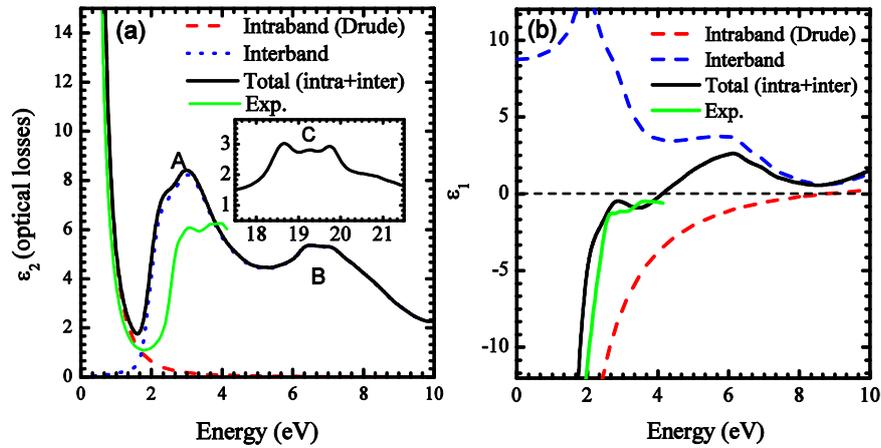

Fig.3 Calculated (a) Imaginary (optical losses) and (b) real permittivity in Au along with the individual contribution from intraband and interband transitions. Inset in Fig (a), shows the imaginary part at higher energy scale to observe other relevant peaks. Green solid line shows the experimental data for comparison.

Here in Fig. 3(a), we present the optical spectra in the energy range up to 10 eV (see inset for high energy scale). One can see the interband onset at ~ 1.8 eV followed by other major peaks at 3 eV and 7 eV. Apart from this, there are some more peaks at high energy scale ~19-20 eV as shown in the inset of Fig 3(a) (indicated as C point). Our presented spectra agree well with the reported calculations for the energy range 0-25 eV [25]. One can interpret the losses shown in Fig. 3 (a) from interband and intraband contribution. The interband losses in Au are high above 1.8 eV, on the other hand, intraband losses are negligible above 2 eV and very high in low energy (NIR) region. Overall, the essential features in the permittivity of Au are captured well in our work and consistent with experiments [36] and theory [25], which confirms the reliability of adopted methodology.

Figures 4(a) and 4(b) show the calculated real and imaginary parts of dielectric functions of $Ti_{1-x}Zr_xN$ alloy system. It is noticeable that TiN shows more loss compared to ZrN and this loss decreases with Zr alloying in $Ti_{1-x}Zr_xN$ alloy. The optical losses in these compounds can be correlated to the electronic structure. The position of $p$ states are different for TiN and ZrN (c.f. Fig. 2), which generally influence the $\omega_{int}$ or optical loss. For example in the case of TiN, N $p$ states are below ~1.9 eV from $E_f$ whereas for ZrN, these states are well below 2.6 eV from $E_f$. As discusses in the previous section, this is the reason that the threshold energy for interband transitions is 1.9 eV for TiN and 2.6 eV for ZrN. The calculated optical parameters such as plasma frequency ($\omega_p$), cross-over frequency ($\omega_c$: corresponding value when $\varepsilon_1(\omega)$ crosses zero), width of plasmon (WP: corresponding value of $\varepsilon_2(\omega)$ when $\varepsilon_1(\omega) = 0$), and onset of interband transitions ($\omega_{int}$) are shown in Table 1, which are in good agreement with the reported experimental data for TiN and ZrN [37-40]. In addition, our calculated $\omega_p$ value of 7.49 eV for TiN and 8.63 eV for ZrN, are in good agreement with the previously reported DFT values of 7.62 eV and 8.82 eV for TiN and ZrN, respectively [27].

Figure 4(c) shows the calculated frequency dependent energy loss function of $Ti_{1-x}Zr_xN$ alloy. The loss function has peaks ($ELS_{peak}$) at the frequencies where the real part of dielectric function becomes zero. Therefore one can verify the $\omega_c$ with these peak maxima of energy loss function. Estimated energies of the low-energy $ELS_{peak}$ of bulk plasmon from loss function are depicted in Table 1. The peak for TiN is at 2.4 eV and a blue shift in peaks are observed for $Ti_{1-x}Zr_xN$ with increasing Zr concentration. Calculated low-energy plasma excitation of TiN at 2.4 eV and ZrN at 3.3 eV, are in reasonable agreement with measured peak at 2.8 eV [39] and 3.6 eV [40], respectively. However, one should note that local-filed effect, spin-orbit coupling effect and quasi particle (GW) effect may lead to minor deviations of ± 0.5 eV in reported values, which however was not considered here.

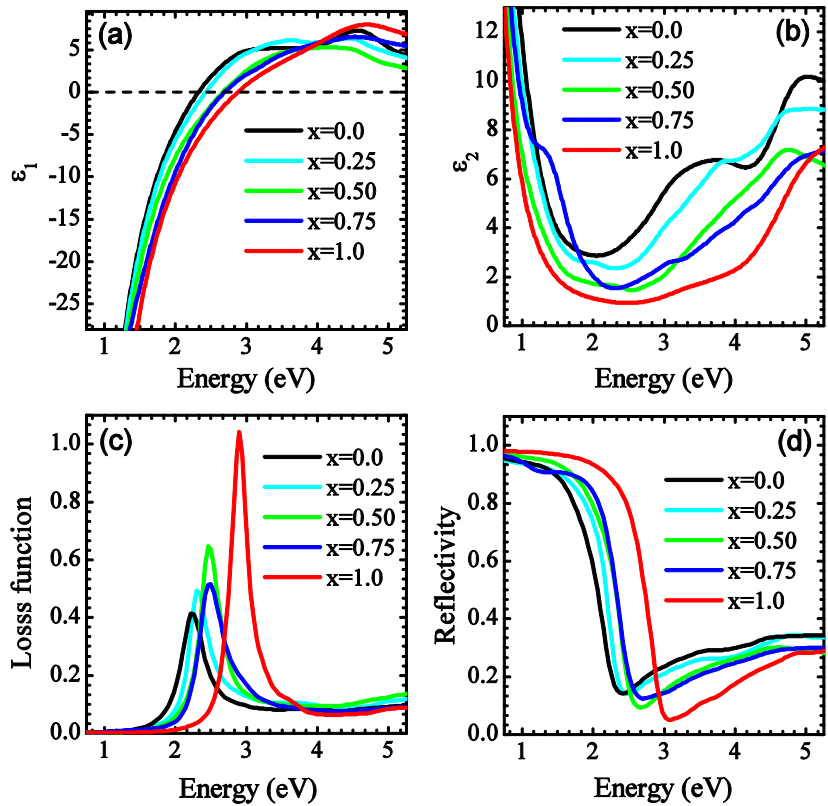

Fig.4 Calculated (a) real permittivity, (b) imaginary permittivity, (c) loss function, and (d) normal incident reflectivity of $Ti_{1-x}Zr_xN$ (x = 0, 0.25, 0.5, 0.75 and 1) alloy system.

Figure 4(d) shows the calculated reflectivity of alloy system at normal incident. Qualitatively, one can estimate the appearance of these materials in bulk based on the calculated reflectivity. Larger reflectivity (>60%) suggests the metallic like surface for these alloys in low wavelength region. TiN shows large reflectivity than ZrN. However, at longer wavelength, the trend is opposite and ZrN shows larger reflectivity than TiN. Note that a reflectivity edge is correlated to the cross-over frequency. Our analysis shows that the bulk plasmon energy of $Ti_{1-x}Zr_xN$ alloy falls in the range of visible wavelength and one can tune this energy by varying Ti/Zr concentration accordingly in the range of 2.3 to 2.9 eV (430 to 540 nm in wavelength).

Overall, one can summarize that the position of $E_f$, nitrogen $p$ states and band edge energies at Γ point influence the optical loss in low energy region for these materials and there is basically monotonic shifts. As a result, one can tune the properties of the alloy system by varying Ti or Zr composition. For example it is possible to have a material having arbitrary $\omega_c$ between 2.32 eV and 2.86 eV by tuning the alloy composition. Monotonic changes in the optical properties indicate that the extreme cases are found either at x = 0 (TiN) or x = 1 (ZrN). Due to the low onset of the interband transitions, TiN have highest loss in low frequency region of the visible spectrum and can be used for applications in broadband light absorption as discussed in the next section. In contrast, ZrN has the lowest loss and can be a good choice for plasmonic waveguides and sensing applications.

Table 1. Calculated optical parameters, plasma frequency ($\omega_p$), cross-over frequency ($\omega_c$), width of plasmon (WP), onset of interband transitions ($\omega_{int}$), and low-energy bulk plasmon peak (ELS$_{peak}$) of Ti$_{1-x}$Zr$_x$N alloy system along with Au for comparison.

| Material | $\omega_p$ (eV) | $\omega_c$ (eV) | WP (eV) | $\omega_{int}$ (eV) | ELS$_{peak}$ (eV) |
|---|---|---|---|---|---|
| TiN | 7.49 (6.69-8.7)$^{a,b}$ | 2.32 | 3.12 | 1.95 | 2.38 (2.8)$^c$ |
| Ti$_{0.75}$Zr$_{0.25}$N | 7.48 | 2.43 | 2.41 | | 2.41 |
| Ti$_{0.5}$Zr$_{0.5}$N | 7.70 | 2.67 | 1.54 | 2.25 | 2.65 |
| Ti$_{0.25}$Zr$_{075}$N | 7.79 | 2.71 | 1.95 | | 2.68 |
| ZrN | 8.63 (7.17-8.02)$^{a,b}$ | 2.86 | 1.09 | 2.60 | 3.26 (3.6)$^d$ |
| Au | 8.76 (8.9)$^e$ | 4.12 | 5.49 | 1.65 (2.3)$^e$ | 4.11 |

$^a$From Phys. Scr. 25, 775 (1982); $^b$From Adv. Mat. 25, 3264 (2013); $^c$From Phys. Rev. B 30, 1155 (1984); $^d$From Phys. Rev. B 31, 1244 (1985) ; $^e$From Laser Photon. Rev. 4, 795-808 (2010)

## 4. Discussion

Here we discuss the performances of TiN, ZrN and their alloys Ti$_x$Zr$_{1-x}$N with Au and Ag in terms of various quality factors. To visualize the performance of these materials for different classes of plasmonic devices, we apply two quality factors defined for localized surface plasmon resonance (LSPR) and surface plasmon polariton (SPP) which are localized and propagating plasmonic oscillations at the surface of metallic components, respectively. Note that there are already some experimental demonstration on both LSPR [42, 43] and SPP [44] with TiN. Their quality factors, Q$_{LSPR}$ and Q$_{SPP}$, can be expressed in terms of their dielectric functional formulism as shown in Eqs. (5) and (6), respectively [5],

$$Q_{LSPR} = \frac{\omega \frac{d\varepsilon_1(\omega)}{d(\omega)}}{2\varepsilon_2(\omega)} \quad (5)$$

$$Q_{SPP} = \frac{\varepsilon_1(\omega)^2}{\varepsilon_2(\omega)} \quad (6)$$

Calculated Q$_{LSPR}$ and Q$_{SPP}$ are shown in Figs. 5(a) and 5(b), respectively, along with those of Au and Ag for comparison.

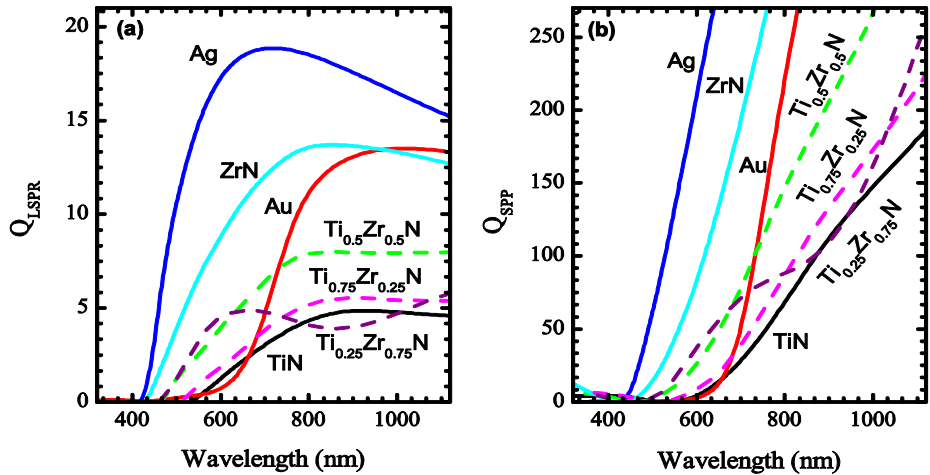

Fig.5 Comparison of (a) Q$_{LSPR}$ and (b) Q$_{SPP}$ of Ti$_{1-x}$Zr$_x$N alloy system with those of Au (red line) and Ag (blue line).

As seen from above figures, Ag outperforms other materials in the whole visible spectrum. ZrN is the second best material followed by Au. Our results are in reasonable agreement with the experimental studies [12, 13], where $Q_{LSPR}$ of epitaxial ZrN film shows better performance than Au beyond $\lambda = 400$ nm. It is worth to mention that in our calculations, we assumed perfect crystal which are free from defects and grain boundaries. Thus in other works, we consider ideal cases which show the best possible performances for each material. In real samples, there are many defects and grain boundaries which can deteriorate the optical properties substantially. Hence, considering the chemical stability and cost of ZrN, the results suggest that ZrN is a practically appealing for LSPR and SPP applications in visible range which include plasmon-assisted bio-imaging, surface enhanced Raman scattering and plasmonic waveguides.

Among the applications of surface plasmon, the LSPRs excited at plasmonic nanostructures have found applications which utilize both scattering and absorption properties. To study the LSPRs of $Ti_{1-x}Zr_xN$ alloy system, a simple case where a nanosphere embedded in an isotropic homogenous host material is considered and the scattering ($Q_{sca}$) and absorption ($Q_{abs}$) efficiencies are calculated by using the numerical solution of Mie theory [41]. Using the DFT calculated bulk dielectric functions shown in Fig. 4(a) and 4(b) and taking the radius and host index of 50 nm and 1.33, respectively, we plotted the $Q_{sca}$ and $Q_{abs}$ efficiencies of $Ti_{1-x}Zr_xN$ alloy system along with those of Au and Ag as shown in Figs. 6(a) and 6(b). One can notice that barring Ag, ZrN has the highest $Q_{sca}$ value of 6.3 at 559 nm (2.22 eV). Similarly, TiN has broader resonance with a maximum $Q_{sca}$ value of 2.36 at 652 nm (1.90 eV). Absorption efficiency of these materials, on the other hand, shows opposite trend as TiN has the highest $Q_{abs}$ value of 3.34 at 634 nm (1.95 eV) which is slightly better than that of Au. Overall, our computational analysis shows that absorption efficiency of TiN is better than Au while scattering efficiency, on the other hand, is not superior which however is in good agreement with experimental finding by Guler et. al. [43].

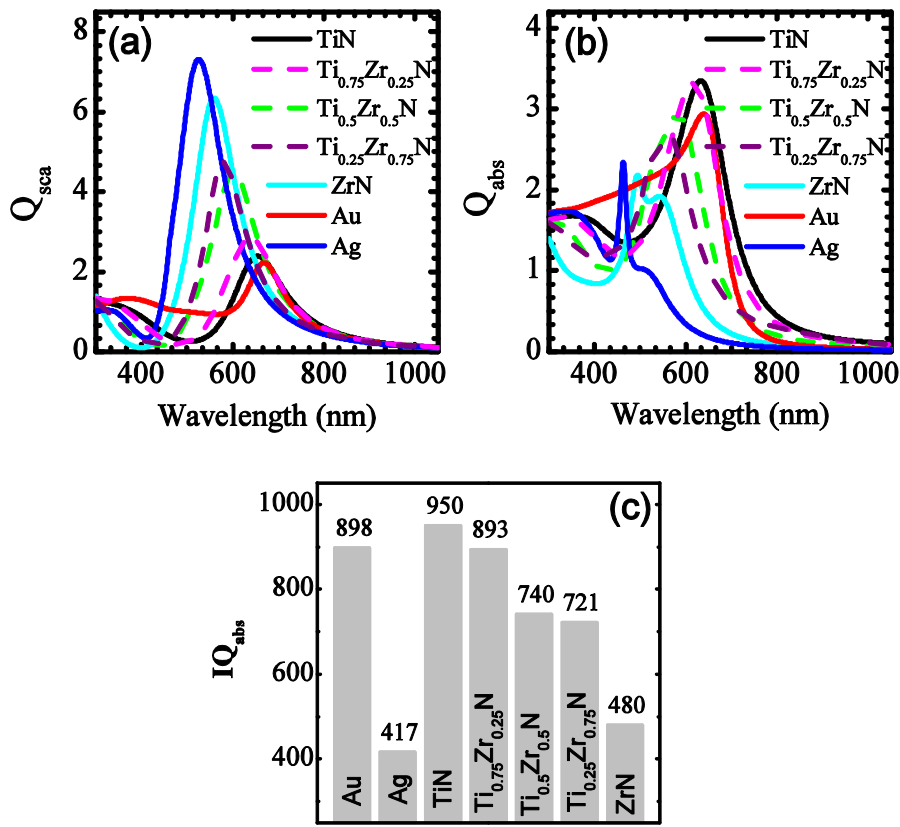

Fig.6 Comparison of (a) scattering, (b) absorption and (c) integrated absorption efficiencies of $Ti_{1-x}Zr_xN$ alloy system with those of Au and Ag.

While the value of $Q_{abs}$ is important at single wavelength illumination, the integrated absorption efficiency is a figure of merit for broadband absorption. With this regard, we calculate the integral of $Q_{abs}$ ($IQ_{abs}$) at 300-1300 nm, which is the transparent region of water, and plotted in Fig. 6(c). It is observed that $IQ_{abs}$ of TiN (950) and $Ti_{0.75}Zr_{0.25}N$ (893) are higher or comparable, respectively than that of Au (898). This comparison shows that TiN and the alloy nanoparticles can be good broadband absorbers [42-44].

## 5. Conclusion

Ternary plasmonic compound $Ti_{1-x}Zr_xN$ was studied to examine the correlation between electronic band structure and optical properties, as well as the tunability of plasmonic properties through the band engineering. Our detailed investigation based on first-principles DFT calculations and electromagnetic simulations (Mie theory) revealed that $Ti_{1-x}Zr_xN$ alloy system possess a great potential to be used as tunable plasmonic materials with their highly metallic properties as well as low optical losses. Among the $Ti_{1-x}Zr_xN$ alloys, ZrN is found to be the best low-loss material whereas TiN has higher loss than ZrN. This was due to the low threshold onset of interband energy (1.95 eV) than ZrN where threshold energy for onset of interband transitions is higher (2.60 eV). Due to isostructural nature of these alloy systems, a monotonic behaviors in electronic and optical properties were observed with cation alloying. Such monotonic property is beneficial if one wants to find a plasmonic material with a specific cross-over wavelength. We believe that the alloy system of $Ti_{1-x}Zr_xN$ further broaden the choices of alternative plasmonic materials.

## Acknowledgments
M.K. would like to thanks Dr. Byungki Ryu of Korea Electrotechnology Research Institute for useful discussion. This work was supported by the Core Research for Evolutional Science and Technology (CREST) program from the Japan Science and Technology Agency (JST). It was also supported by the JSPS KAKENHI Grant Number 15K17447.